\documentclass[aip,rsi,reprint,graphicx]{revtex4-1} 

\usepackage{graphicx}
\usepackage{amsmath, amsthm, amssymb}
\usepackage{float}

\draft 

\begin{document}


\title{Onion-peeling Inversion of Stellarator Images} 
\thanks{Contributed paper published as part of the Proceedings of the 21st 
        Topical Conference on High-Temperature Plasma Diagnostics, Madison, 
        Wisconsin, June 2016.\\}



\author{K.C.~Hammond}
\email{kch2124@columbia.edu}
\affiliation{Department of Applied Physics and Applied Mathematics, 
Columbia University, New York, NY 10027}

\author{R.~Diaz-Pacheco}
\affiliation{Department of Applied Physics and Applied Mathematics, 
Columbia University, New York, NY 10027}

\author{Y.~Kornbluth}
\affiliation{Yeshiva University, New York, NY 10033}
\affiliation{Present Address: Massachussets Institute of Technology, Cambridge, MA 02139}

\author{F.A.~Volpe}
\affiliation{Department of Applied Physics and Applied Mathematics, 
Columbia University, New York, NY 10027}

\author{Y.~Wei}
\affiliation{Department of Applied Physics and Applied Mathematics, 
Columbia University, New York, NY 10027}



\date{\today}

\begin{abstract}
An onion-peeling technique is developed for inferring the emissivity
profile of a stellarator plasma from a two-dimensional image acquired
through a CCD or CMOS camera. Each pixel in the image is treated as an
integral of emission along a particular line-of-sight. Additionally,
the flux surfaces in the plasma are partitioned into discrete layers,
each of which is assumed to have uniform emissivity. If the topology
of the flux surfaces is known, this construction permits the
development of a system of linear equations that can be solved for the
emissivity of each layer. We present initial results of this method
applied to wide-angle visible images of the CNT stellarator
plasma. 
\end{abstract}

\pacs{}

\maketitle 

\section{Introduction} 

A number of well-established techniques exist for inferring plasma 
emission profiles based on line-of-sight measurements.
The measurements are interpreted as line integrals of emission and 
are typically acquired in 1D or 2D arrays.
This poses a question of ``inverting'' the acquired datasets  
into 2D or 3D maps, respectively, of plasma emissivity of the particle or 
wavelength of interest. 

Tomography has succeeded at this 
in tokamaks \cite{Ingesson} and stellarators \cite{Sallander}, 
primarily for X-ray and visible emission \cite{Gota,Asai}.
Tomography does not require knowledge of the flux surface geometry, 
although the geometry can constrain the tomography for better results.
A disadvantage, though, is that it requires several cameras around the plasma. 

Under the assumption of the flux surface geometry being perfectly 
known, a single camera is sufficient. If, additionally, the flux 
surfaces are axisymmetric, one can use the Abel transform \cite{Abel,Dasch}. 
This is appropriate in quiescent tokamak and spherical tokamak plasmas. 

In stellarators it is reasonable to assume a good knowledge of the 
flux surfaces, which are nearly entirely determined by external 
currents and are magnetohydrodynamically more quiescent 
than in tokamaks and sperical tokamaks. This is especially true when 
the ratio of kinetic pressure to magnetic pressure, $\beta$, is low: in that 
case, stellarator equilibria are easier to compute, faithful 
to experiments and magnetohydrodynamically stable. 
However, stellarator flux-surfaces are obviously not axisymmetric. 

The onion peeling algorithm \cite{Dasch} 
can be considered a generalization of the Abel inversion to non-axisymmetric 
problems. Onion peeling is used here to invert wide-angle 
visible images of a stellarator plasma for the first time. 
The work was performed at the CNT stellarator \cite{Pedersen} 
and takes advantage 
of a recent experimental and numerical study of its field errors \cite{Ken}, 
giving good confidence in the knowledge of its flux surfaces. 
The method is described in Sec.\ref{Sec:Met}. 
Sec.\ref{Sec:Fwd} is devoted to the ``forward problem'': toy models of the 
emissivity profile are 
translated in the corresponding images expected to be acquired by the camera. 
The synthetic image corresponding to an edge-peaked profile is in best  
qualitative agreement with the actual experimental images. 
The inverse problem is solved in Sec.\ref{Sec:ExpRes}. The method is first
tested with a glow discharge plasma for validation and then applied to 
a microwave-heated plasma.

\section{Method}\label{Sec:Met}

\begin{figure}
    \begin{center}
    \includegraphics[width=0.32\textwidth]{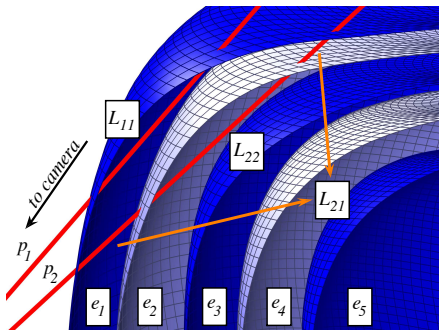}
    \end{center}
    \caption{Schematic illustration of the onion-peeling method.
             The red lines represent lines-of-sight of two camera pixels
             whose brightness is denoted by $p_1$ and $p_2$. The lines-of-sight
             both travel through one or more layers of plasma, whose borders
             are shown as alternating blue and white surfaces. The lengths
             of the segments within the respective surfaces constitute the 
             components of the $L$ matrix as defined in the text. Note that
             some components (like $L_{21}$ here) are actually due to the sum
             of two or more segments due to multiple crossings of a layer.
             The terms $e_i$ are the emissivities of each layer.}
    \label{fig:schematic}
\end{figure}

The method used in this paper for reconstructing the emissivity profile 
relies on two principal assumptions. The first is that the plasma can be 
modeled as a set of nested discrete layers, each of which has a uniform 
emissivity. In the context of a toroidal magnetic confinement device, these
layers are bounded by flux surfaces (Fig.~\ref{fig:schematic}). The second
assumption is that the emissive layers contribute to the brightness of each 
camera pixel in proportion to the distance that 
the pixel's line-of-sight travels through each layer.

With these assumptions in place, the brightness $p$ of each pixel is related
to the emissivity of each layer by a system of linear equations,

\begin{equation}
    b\mathbf{p} = L\mathbf{e},
\label{eqn:fwd}
\end{equation}

\noindent where $p_i$ is brightness of the $i^{th}$ pixel, $b$ is a constant of
proportionality, $L_{ij}$ is the total distance traveled by the $i^{th}$ pixel's
line-of-sight throught the $j^{th}$ layer, and $e_j$ is the emissivity of the 
$j^{th}$ layer. The elements of the matrix $L$ can be calculated based on 
knowledge of (1) the flux surfaces, (2) the camera's position, 
orientation, and field of view, and (3) the positions of any obstacles
(such as CNT's in-vessel coils) obstructing the camera's view of portions of 
the plasma. 
For the low-$\beta$ plasmas studied in CNT for
this work, any differences from the measured vacuum flux surfaces were 
neglected.

The pixel data tends to be noisy and can exhibit nontrivial correlations. 
Hence, if one is to reconstruct the emissivity
profile based on pixel data, it is best to have many more pixels than layers
to reconstruct. An unbiased estimator of the emissivity profile can then be
expressed as\cite{jones2006}

\begin{equation}
    \mathbf{e} \approx \left(L^TC^{-1}L\right)^{-1}L^TC^{-1}~b\mathbf{p},
\label{eqn:inv}
\end{equation}

\noindent where $C$ is the covariance matrix for the pixel noise. 

When the emissivity profile $\mathbf{e}$ is calculated in this way, it is 
informative to plug it back into Eq.~\ref{eqn:fwd} to synthesize an image 
$\mathbf{p}^*$ for comparison with the original image.

\section{Forward problem}\label{Sec:Fwd}

\begin{figure}
    \includegraphics[width=0.25\textwidth]{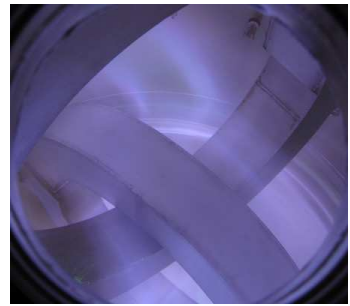}
    \caption{Photograph of a typical argon plasma in CNT heated with 1 kW 
             ECRH with CNT's two in-vessel coils in clear view.}
    \label{fig:argonPhoto}
\end{figure}

As an initial proof-of-concept and a test of the reliability of the 
calculations of $L$, synthetic images were generated based on profiles
that were specified \textit{a priori}. $L$ was calculated to correspond to 
a camera view similar to that of the photograph shown in 
Fig.~\ref{fig:argonPhoto}. The test profiles were given three basic functional 
forms: hollow (linearly increasing from core to edge), uniform, and 
peaked (linearly decreasing from core to edge). 

\begin{figure}
    \includegraphics[width=0.45\textwidth]{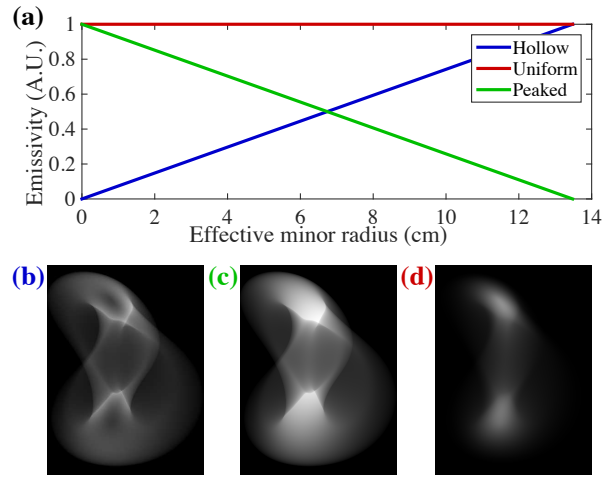}
    \caption{Simulated images of three synthetic profiles.
             (a) profiles; (b) image of hollow profile; (c) image of
             uniform profile; (d) image of peaked profile. The upper halves of
             of (b)-(d) approximate the field of view of the photo in 
             Fig.~\ref{fig:argonPhoto}.}
    \label{fig:forwardProblem}
\end{figure}

These profiles and the resulting synthetic images are shown in 
Fig.~\ref{fig:forwardProblem}. Of the three test profiles, the hollow one
(Fig.~\ref{fig:forwardProblem}b) has the best qualitative agreement with 
the photograph in Fig.~\ref{fig:argonPhoto}, suggesting that the plasma in 
the photo (which is typical of the $\approx 1 kW$ electron cyclotron resonant
heating (ECRH) discharges studied in CNT)
has greater visible emission at the edge than in the core. This is consistent
with a higher rate of recombination reactions occurring in the colder edge.

\section{Inversion of experimental images}\label{Sec:ExpRes}

\subsection{Image processing}
\label{subsect:imProc}

\begin{figure*}
    \begin{minipage}[c]{0.65\textwidth}
        \includegraphics[width=0.95\textwidth]{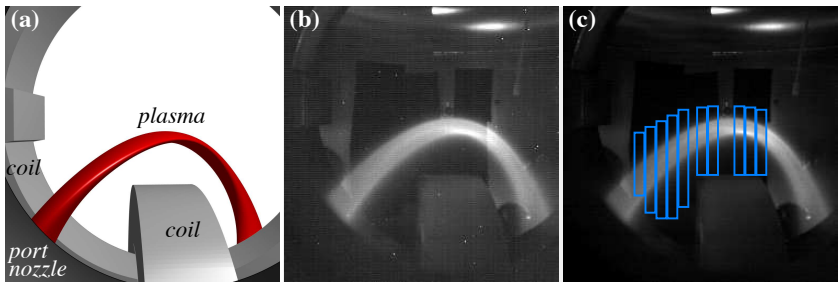}
    \end{minipage}\hfill
    \begin{minipage}[c]{0.35\textwidth}
    \caption{(a) Schematic of the key objects in the camera's field of view.
             (b) Raw image of a typical glow discharge 
             (Sect.~\ref{subsect:glows}).
             (c) The same image, after noise subtraction, with boxes 
             indicating the regions where pixel data were used for 
             reconstructions (Sect.~\ref{subsect:imProc}).}
    \label{fig:bboxes}
    \end{minipage}
\end{figure*}

Images used for profile reconstructions were acquired by a high-speed 
CMOS camera manufactured by Canadian Photonics. The camera was placed outside
the vacuum chamber with a view through a fused silica window. The field of 
view is shown in Fig.~\ref{fig:bboxes}. As 
Eq.~\ref{eqn:fwd} does not account for external sources of light, 
sheets of low-outgassing black foil manufactured by Acktar were fixed to 
the internal vessel walls to prevent reflection in the camera's lines-of-sight.

To account for pixel noise, a large number of frames (greater than or equal
to the total number of pixels to be analyzed) was acquired with no light sources
present. The mean value of the noise acquired for each pixel was then
subtracted from the corresponding pixel in a plasma image. The result of this
subtraction is shown in Fig.~\ref{fig:bboxes}b. The background frames were 
also used to compute the pixel noise covariance matrix $C$ for the 
inversion (Eq.~\ref{eqn:inv}). Optical noise originating from the chamber was
neglected. The propagated error $\sigma_{\text{noise},j}$ for the $j^{th}$ 
layer in the profile was calculated as the square root of the $j^{th}$ 
diagonal element of the posterior covariance matrix\cite{jones2006}, given
by $\left(L^TC^{-1}L\right)^{-1}$.

For each plasma image, ten independent inversions were conducted 
using disjoint subsets of the pixels (shown as blue
boxes in Fig.~\ref{fig:bboxes}b). The subsets used in this work contained 
between 560 and 860 pixels. To be included in a subset, a pixel's 
line-of-sight needed to terminate against black foil. The ten different 
profiles $\mathbf{e}$ obtained from each of the subsets were then averaged 
to obtain a final reconstructed profile. This was done to partially cancel
the effects
of small errors in the camera alignment, which would otherwise cause some 
contributions to some pixels to be misattributed to layers not within their
lines-of-sight. The standard deviation among these measurements will be 
notated as $\sigma_\text{align}$.
 
The total uncertainty of the emission from the $j^{th}$ layer was then 
computed from the quadrature sum of the alignment error $\sigma_\text{align}$
and the noise error $\sigma_\text{noise}$ averaged over $N$ pixel subsets:

\begin{equation}
    \sigma_j = 
        \sqrt{\frac{{\sigma_{\text{align},j}}^2}{N} + 
              \left(\sum_n^N\frac{{\sigma_{\text{noise},nj}}}{N}\right)^2}
\end{equation}

\subsection{Reconstructions of glow discharges}
\label{subsect:glows}

\begin{figure}
    \includegraphics[width=0.45\textwidth]{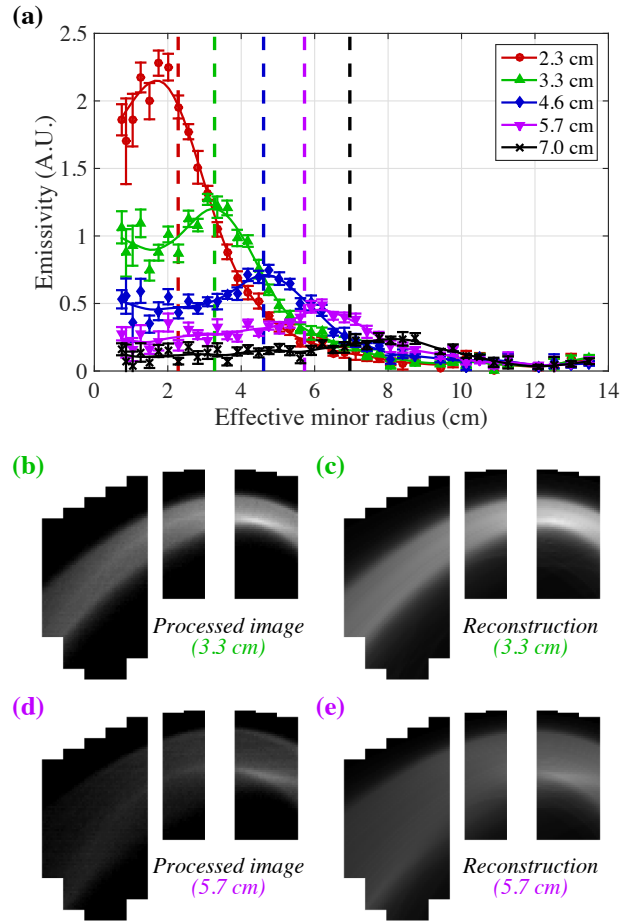}
    \caption{Reconstructed profiles and images of glow discharges. 
             (a) Profiles with the emissive filament located on six different
             flux surfaces with effective minor radii given in the legend and
             as vertical dashed lines.
             Solid curves are fits to cubic smoothing splines, which
             are used for the image reconstructions. 
             (b)-(c) Image and reconstruction for the 3.3 cm filament position 
             in the regions where pixel data were used for reconstructions 
             (Fig.~\ref{fig:bboxes}b).
             (d)-(e) Image and reconstruction for the 5.7 cm filament position.}
    \label{fig:reconstr_glow}
\end{figure}

Field line and flux surface visualizations are often utilized in CNT for 
diagnostic alignment. \cite{brenner2008} These are discharges maintained by
a heated, biased filament. They tend to emit a visible glow that 
is localized to the flux surface where the filament is located. In the 
case of rational or near-rational surfaces, the glow is restricted to the flux
tube connecting one side of the filament to the other. On non-rational 
surfaces, however, the entire surface tends to be emissive. 

While this glow is not perfectly uniform, it is nonetheless expected that an 
inversion of such a glow would result in an emission profile that is 
peaked around the layer where the emitter is located. Hence, inversions
of glow discharges can serve as tests of the method's validity.

\begin{figure*}
    \begin{minipage}[c]{0.75\textwidth}
        \includegraphics[width=\textwidth]{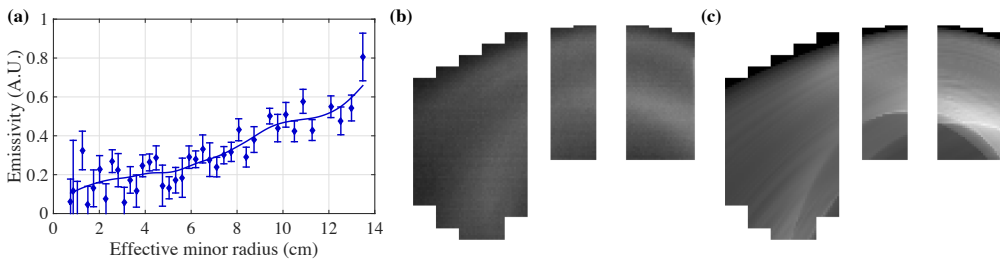}
    \end{minipage}\hfill
    \begin{minipage}[c]{0.25\textwidth}
    \caption{Reconstruction of a 1 kW ECRH discharge. (a) Emissivity profile
             with a cubic spline fit. (b) Photographic data used for the 
             reconstruction. (c) Synthesized image based on the reconstructed
             profile.}
    \label{fig:reconstr_uwave}
    \end{minipage}
\end{figure*}

Quiescent glow discharges were filmed at 25 frames per second with 15 ms
exposures at the camera's lowest gain setting. 
Fig.~\ref{fig:reconstr_glow}a 
shows emissivity profiles from glow discharges
from several filament locations at successively larger effective minor
radii. As expected, each profile is peaked. Additionally, the peak locations 
agree well with the locations of the filament, which were measured 
independently through field line mapping procedures similar to those described
in Refs.~\cite{pedersen2006,jaenicke1993}.
Glow discharges on surfaces with larger minor radii tend to have lower 
calculated emissivity overall; this is consistent with visual inspection of 
the discharges.

Fig.~\ref{fig:reconstr_glow}b-e show raw and reconstructed images corresponding
to two of the profiles from Fig.~\ref{fig:reconstr_glow}a. While the agreement
is qualitatively good in many respects, there are some features from the 
photos that are not replicated in the reconstructions. Perhaps the most 
noticeable are the bright, narrow streaks that are visible in the middle of the 
plasmas 
in the center and left of Figs.~\ref{fig:reconstr_glow}b,d. The
streak is a nonuniformity on its surface (even on non-rational surfaces,
regions with short connection lengths to the filament tend to glow brigher than
the rest of the surface). Such a nonuniformity is not expected to carry over
in the inversion process, due to the assumption of uniform emissivity within
a layer discussed in Sec.~\ref{Sec:Met}.

\subsection{Reconstruction of an ECRH discharge}

1 kW ECRH discharges lasting for roughly 7 ms were filmed at 500 frames per 
second with 2 ms exposures at the lowest gain setting. 
Fig.~\ref{fig:reconstr_uwave} shows a profile, image data, and reconstructed
image from a typical frame. Note that, except for the outermost layer,
the profile follows a roughly linear trend similar to the hollow test profile
in Fig.~\ref{fig:forwardProblem}. In contrast to the image comparisons from 
Fig.~\ref{fig:reconstr_glow}, in which the image data had fine structure that
did not appear in the reconstructions, the reconstruction for the ECRH
discharge in Fig.~\ref{fig:reconstr_uwave}c has fine structure that is not 
visible in the image data (Fig.~\ref{fig:reconstr_uwave}b). Much of the 
structure in the reconstruction results from the jump in emissivity at the 
edge as seen in the 
profile, which would result in bright spots wherever the outermost layer
is nearly tangent to lines-of-sight. It is not fully understood why the 
jump in emissivity appears in the calculated profile despite not appearing
experimentally. It is 
conjectured that it may be a spurious effect resulting from emission outside
the closed flux surfaces; \textit{i.e.}, in the scrape-off layer.

\section{Summary, Conclusions and Future work}

In summary, a method has been implemented for deducing the emissivity profile
of a nonaxisymmetric toroidal plasma based on images from a single camera
location.
Reconstructions of flux surface visualizations yielded peaked profiles as 
expected, thereby serving as a promising test of concept. The fact that 
the non-uniform features from the photographs of the glow discharges did not 
appear
in the reconstructed images serves to emphasize that this method will not
account for non-uniformities in emission across a layer. A reconstruction of
an ECRH discharge contained some unexpected features but still exhibited 
an overall hollow profile, consistent with expectations.

Future work will include incorporating the 
scrape-off layer in the onion-peeling inversion. 
A further improvement would be to cover larger portions of the CNT vessel, 
as well as the interlocked coils, with the light-absorbing material. 
As a result, larger portions of the experimental images would be suitable for 
inversion. 

The technique will then be deployed to study
the variation of the emissivity profile over the course of ECRH heating 
pulses. Profiles will be compared under different magnetic configurations
and heating locations. Optical filters may also be placed in front of the 
camera lens to isolate particular emission lines.

\begin{acknowledgments}

The authors would like to thank the PPPL University Collaboration Program
for the donation of the fast camera. 
This work was funded internally by Columbia University.

\end{acknowledgments}

\end{document}